\documentstyle[prl,aps,multicol,epsf]{revtex}
\begin{document}
\draft
\title{Relative dispersion in fully developed turbulence:\\
from Eulerian to Lagrangian statistics in synthetic flows}

\author{G. Boffetta $^{1,2,3}$, A. Celani $^{4,2,3}$, A. Crisanti $^{5,6}$
	and A. Vulpiani $^{5,6}$}
\address{$^1$Dipartimento di Fisica Generale, Universit\`a di Torino,
         Via Pietro Giuria 1, 10125 Torino}
\address{$^2$Istituto di Cosmogeofisica, c. Fiume 4, 10133 Torino}
\address{$^3$INFM Unit\`a di Torino Universit\`a, Italy}
\address{$^4$DIAS, Politecnico di Torino, 
	Corso Duca degli Abruzzi 24, 10129 Torino}
\address{$^5$Dipartimento di Fisica, Universit\`a di Roma ``La Sapienza'',
         P.le Aldo Moro 2, 00185 Roma}
\address{$^6$INFM Unit\`a di Roma I, Italy}

\date{\today}

\maketitle

\begin{abstract}
The effect of Eulerian intermittency on the Lagrangian statistics of
relative dispersion in fully developed turbulence is investigated.
A scaling range spanning many decades is achieved by generating a 
multi-affine synthetic velocity field with prescribed intermittency 
features.
The scaling laws for the Lagrangian statistics are found to depend
on  Eulerian intermittency in agreement with a multifractal
description. As a consequence of the Kolmogorov's law,
the Richardson's law for the variance of pair separation is not affected 
by intermittency corrections.
\end{abstract}

\begin{multicols}{2}


Understanding the statistics of particle pairs dispersion 
in turbulent velocity fields is of great interest for both theoretical and 
practical implications. 
Since fully developed turbulence displays well known, non-trivial 
universal features in the Eulerian statistics of velocity differences 
\cite{MY75,Frisch95}, it represents a starting point for the investigation
of the general problem of the relationship between Eulerian
and Lagrangian characteristics.

Since the pioneering work by Richardson \cite{Richardson26},
many efforts have been done to confirm experimentally \cite{MY75}
or numerically \cite{ZB94,EM96,FV98} his law. 
Most of the previous works concerning the validation of the 
Richardson's law have been focused mainly on the numerical prefactor 
(Richardson constant \cite{MY75}). 
Also theoretically there are very few attempts to investigate possible
corrections stemming from Eulerian intermittency 
\cite{Novikov89,CPV87,GP84} (see also \cite{PC90} and \cite{BS94}
for a discussion on single-particle Lagrangian intermittency).

This is quite surprising compared with the enormous amount 
of literature concerning the intermittency correction for the Eulerian
statistics \cite{Frisch95,BJPV98}. The main obstacle to a deeper 
investigation of relative dispersion is essentially the lack of sufficient 
statistics due technical difficulties in laboratory experiments and to 
the moderate inertial range reached in direct numerical simulations.

In this Letter we present a detailed investigation of the
statistics of relative dispersion, obtained by numerical simulations
of the advection of particle pairs on a synthetic turbulent velocity field
with prescribed intermittency features.
Our main result is that there is evidence of intermittency corrections
to Lagrangian scaling laws for relative dispersion,
and these corrections are tightly related to the intermittency of the
Eulerian statistics of velocity differences.

The Richardson's law says that, in fully developed turbulence,
\begin{equation}
\langle R^{2}(t) \rangle \sim t^{3}
\label{eq:1}
\end{equation}
where $R$ is the separation of a particle pair and the average 
is performed over many dispersion experiments or over many particle pairs.
The scaling (\ref{eq:1}) can be obtained by a simple dimensional 
argument \cite{MY75} starting from Kolmogorov's similarity law for 
velocity increments in fully developed turbulence
\begin{equation}
\langle | \delta v^{(E)}(R) | \rangle = 
\langle |{\bf v}({\bf x}+{\bf R})-{\bf v}({\bf x})| \rangle \sim
R^{1/3}
\label{eq:2}
\end{equation}
with $R=|{\bf R}|$.
The pair of  particles separates according to
\begin{equation}
{d {\bf R} \over d t}= \delta {\bf v}^{(L)}({\bf R})
\label{eq:3}
\end{equation}
where $\delta {\bf v}^{(L)}$ represents the velocity difference 
evaluated along the Lagrangian trajectories.
Assuming $\delta v^{(L)}(R) \simeq |\delta v^{(E)}(R) |$ from (\ref{eq:2}) one
obtains $d R^2/dt \sim R \delta v^{(L)}_{\parallel}(R) \sim R^{4/3}$
and hence the Richardson's law (\ref{eq:1}).

To investigate the role of Eulerian intermittency we have performed
extensive numerical investigations of the relative dispersion at very
large Reynolds numbers.  
To accomplish this purpose we have developed a Lagrangian numerical code
for particle pairs whose separation evolves according to (\ref{eq:3}) 
with a realistic turbulent velocity difference.
We consider the Quasi-Lagrangian reference frame \cite{LPP97} moving with
a reference particle. The second particle is advected by the relative 
velocity $\delta {\bf v}({\bf r},t)$ which possesses the same 
single-time statistics of the Eulerian velocity, whenever one considers 
statistically stationary flows. A realistic velocity field
in this reference frame is generating by extending a recently introduced
stochastic algorithm for the generation of multiaffine processes 
\cite{BBCCV98}.
For sake of simplicity we consider, as in \cite{EM96,FV98}, a two 
dimensional velocity field.
The reason is that the relevant aspect for the statistic of particle 
pairs reparation are the scaling laws for the relative velocity 
$\delta{\bf v}({\bf r},t)$, which we take equal
to those of three dimensional turbulence. The extension to a three
dimensional velocity field is not difficult, but more expensive
in terms of numerical resources.

We introduce the stream function $\psi(\bbox{r},t)$ which, in
isotropic conditions, can be decomposed using polar coordinates as
\begin{equation}
\psi(r,\theta,t) = \sum_{i=1}^{N}\sum_{j=1}^{n} 
            \frac{\phi_{i,j}(t)}{k_i} F(k_i r) G_{i,j}(\theta).
\label{eq:2.5}
\end{equation}
Being interested in velocity fields possessing scaling laws on
a large number of decades we use $k_i=2^{i}k_{0}$. The width of
the ``inertial range'' is thus of order $2^{N}$.
The $\phi_{i,j}(t)$ are stochastic processes with characteristic
times $\tau_i=2^{-2i/3}\,\tau_0$, zero mean and 
$\langle |\phi_{i,j}|^p\rangle \sim k_i^{-\zeta_p}$.
An efficient way of to generate $\phi_{i,j}$ is \cite{BBCCV98}:
\begin{equation}
\phi_{i,j}(t) = g_{i,j}(t)\, z_{1,j}(t)\,z_{2,j}(t)\cdots z_{i,j}(t)
\end{equation}
where the $z_{k,j}$ are independent, positive definite, identically
distributed random processes with characteristic time $\tau_k$, while
the $g_{i,j}$ are independent stochastic processes with zero mean,
$\langle g_{i,j}^2\rangle \sim k_i^{-2/3}$ and characteristic time
$\tau_i$.

For a fully developed turbulent velocity field we expect the scaling
$\langle |\psi(r,\theta)|^p \rangle \sim r^{\zeta_p + p}$ which can
be simply achieved by asking that the radial function
$F(x)$ has support only for $x \simeq 1$ and choosing the scaling 
of the random processes $\phi_{i,j}$ as above described. With this
choice the exponents $\zeta_p$ are determined by the probability
distribution of $z_{i,j}$ via
\begin{equation}
\label{eq:zetap}
 \zeta_p = \frac{p}{3} - \log_2\langle z^p\rangle.
\end{equation}
A simple way for constructing the synthetic turbulent field is with 
the following choice:
\begin{equation}
F(x) = x^2(1-x)\ \mbox{\rm for}\ 0\le x \le 1
\end{equation}
and zero otherwise, 
\begin{equation}
G_{i,1}(\theta) = 1, \qquad G_{i,2}(\theta) = \cos(2 \theta + \phi_i)
\end{equation}
and $G_{i,j}=0$ for $j>2$ ($\phi_i$ is a quenched random phase). 
It is worth remarking that this choice is rather general 
because it can be derived from the lowest order expansion for small $r$
of a generic streamfunction in Quasi-Lagrangian coordinates.

The intermittency in the velocity field can be tuned by the set
of parameters entering into the construction of $\phi_{i,j}(t)$. 
In this Letter we shall consider synthetic turbulent fields whose
intermittency corrections to the Kolmogorov scaling, i.e., nonlinear
$\zeta_p$, are close to the true three dimensional turbulence 
exponents \cite{Anselmet84}, i.e. $\zeta_1=0.39$, $\zeta_2=0.72$ 
and so on.

\begin{figure}[ht]
\epsfxsize=220pt\epsfysize=183.68pt\epsfbox{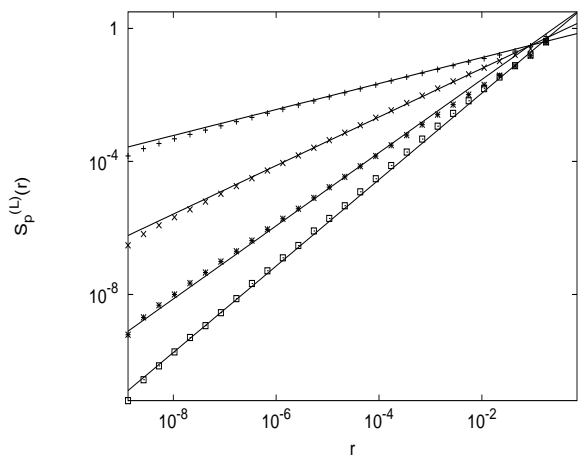}
\caption{Lagrangian longitudinal structure functions $S_{p}^{(L)}(r)$ 
for $p=1,2,3,4$ (from top to bottom) for $N=30$ shells intermittent
velocity field. The continuous lines represent the theoretical scaling with
exponents $\zeta_1=0.39$, $\zeta_2=0.72$, $\zeta_3=1.0$ and $\zeta_4=1.24$.
}
\label{fig1}
\end{figure}

In Figure \ref{fig1} we report the Lagrangian longitudinal structure 
functions 
$S^{(L)}_p(r)=\langle (\delta v^{(L)}_{\parallel}(r))^p \rangle$,
which are computed recording the Lagrangian velocity difference whenever
the pair separation equals $r$. 
Observe the wide inertial range over more than $10$ decades, 
corresponding to an integral Reynolds number $Re \simeq 10^{10}$.
Because of incompressibility the particles separate on average in 
time \cite{Orszag70} and thus also the first order Lagrangian 
structure function is non zero. 
The most interesting and non-trivial result is that scaling exponents for 
the Lagrangian structure functions show up to be exactly the same 
$\zeta_{p}$ of the Eulerian case.
In terms of the multifractal formalism, this result is restated by
saying that the fractal dimension $D(h)$ for the Lagrangian
statistics is the same of the Eulerian one.

With this preliminary results in mind, we can extend the dimensional
argument for the Richardson's law to the intermittent case by using
the multifractal representation.
Following, with a few changes, Novikov \cite{Novikov89} we assume, in the
spirit of the refined similarity hypothesis (RSH) of Kolmogorov, that
$\delta v^{(L)}(R(t)) \sim (\epsilon_{R(t)}\,t)^{1/2}$ and 
$R(t) \sim (\epsilon_{R(t)}\,t^3)^{1/2}$
where $\epsilon_{R}$ is the energy density dissipation at scale $R$. 
Assuming that $\epsilon_{R}\sim \delta v^{(E)}(R)^3/R$ 
and remembering that in the multifractal description of fully 
developed turbulence $\delta v^{(E)}(R)\sim R^h$ with probability 
$P_R(h)\sim R^{3-D(h)}$ \cite{Frisch95,PV87}, a simple calculation leads to
\begin{equation}
\langle R^{p}(t) \rangle \sim 
\int dh\, t^{[3+p-D(h)]/[1-h]}.
\label{eq:4}
\end{equation}
In the limit of time $t$ much smaller than the eddy turnover time at large
scale the integral can be performed by steepest descent method, and
we obtain the scaling laws
$\langle R(t)^p \rangle \simeq t^{\alpha_p}$ where the exponents
are given by
\begin{equation}
\alpha_p = \inf_{h} \left[{p+3-D(h) \over 1-h} \right]
\label{eq:5}
\end{equation}
From the above argument we thus expect that, in general, relative 
dispersion displays anomalous scaling in time (non linear $\alpha_p$).
However there is an interesting result, already obtained in 
\cite{Novikov89}, for the case $p=2$. From the general 
multifractal formalism one has that 
$3-D(h) \ge 1-3h$ and the equality is satisfied for the scaling 
exponent $h_3$ which realizes the third order structure function $\zeta_3=1$. 
From (\ref{eq:5}) follows that $\alpha_2=3$ and thus 
we have that the Richardson's law $\langle R^2 \rangle \sim t^3$ is not affected
by intermittency corrections, while the other moments in general are.
We note that the Lagrangian RSH argument leading to eq. (\ref{eq:4}) is 
just one dimensional reasonable assumption which can be justified only
a posteriori by numerical simulations.
Other different assumptions are possible \cite{CPV87,GP84,Novikov89} leading
to different predictions. 

\begin{figure}[ht]
\epsfxsize=220pt\epsfysize=183.68pt\epsfbox{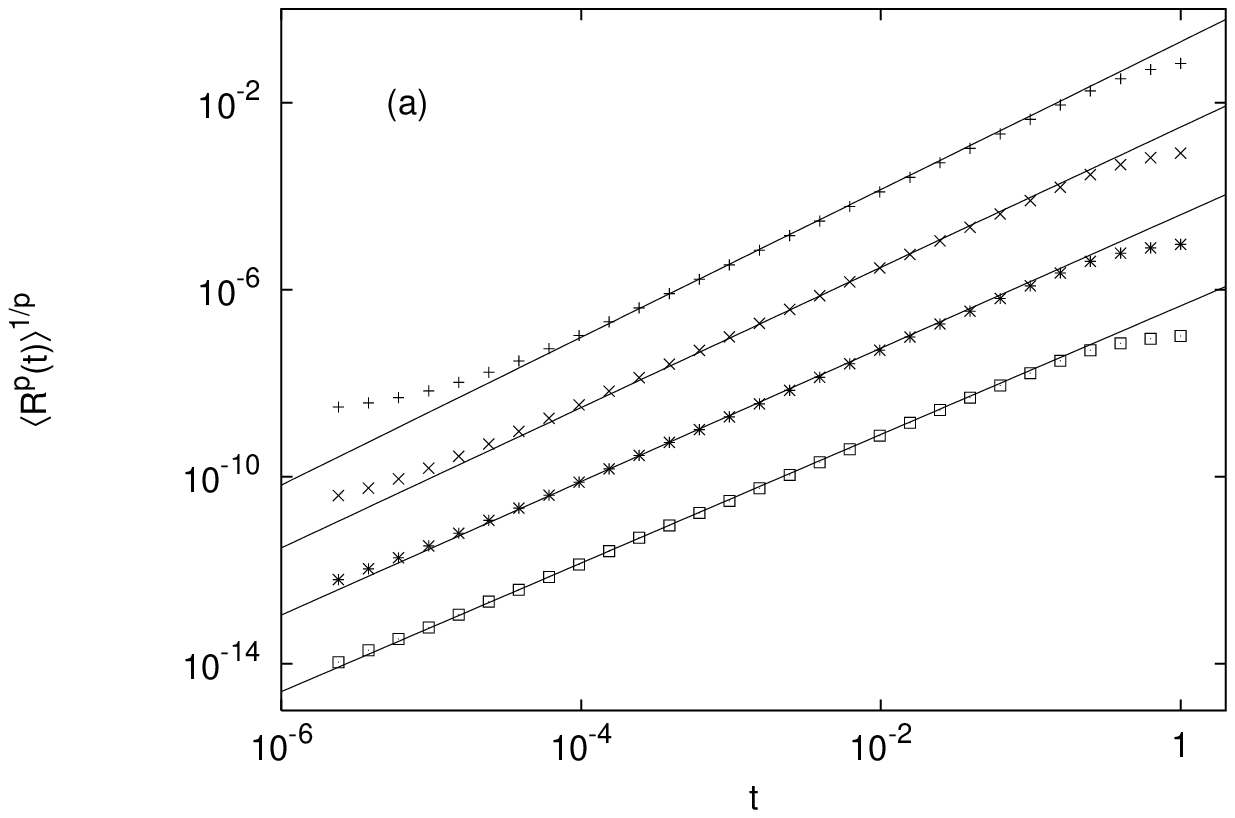}
\epsfxsize=220pt\epsfysize=183.68pt\epsfbox{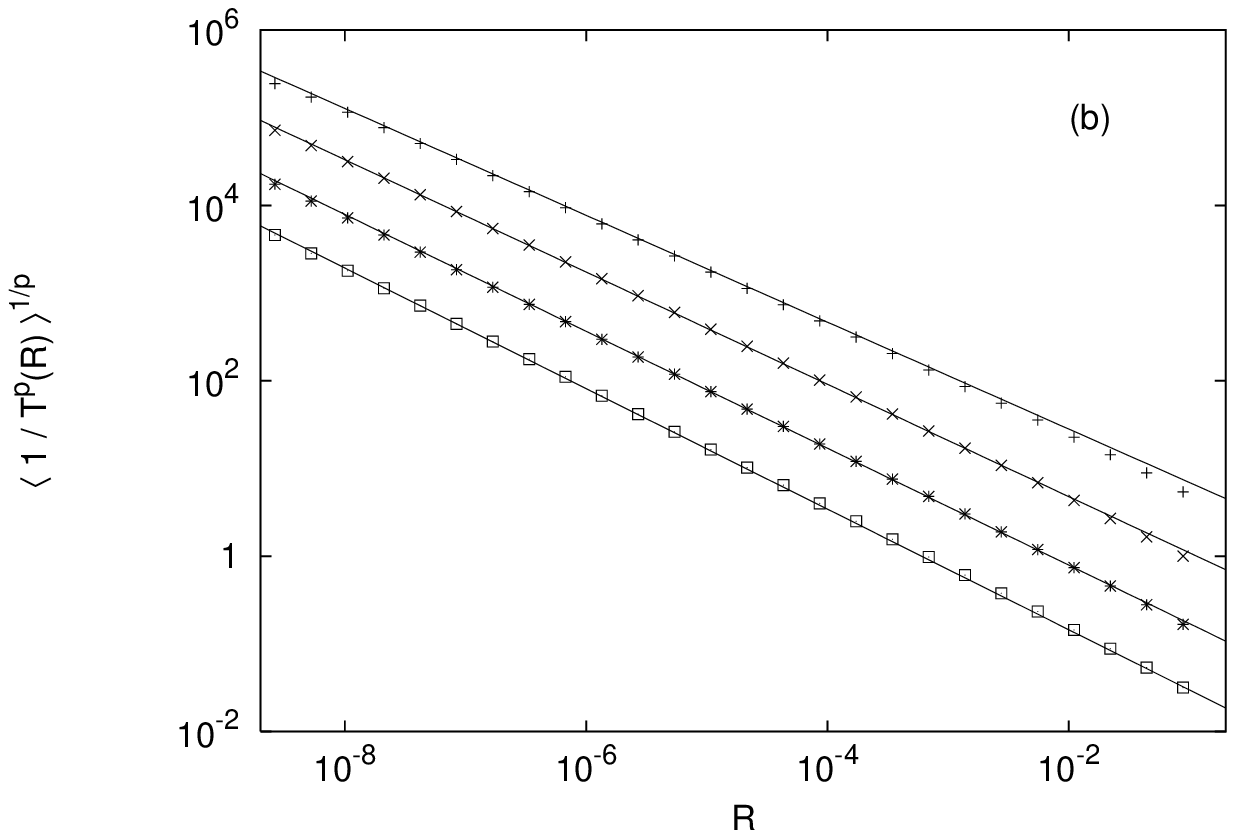}
\caption{Relative dispersion $\langle R^{p}(t) \rangle^{1/p}$ (a)
and inverse doubling times $\langle 1/T^{p}(R) \rangle^{1/p}$ (b)
for $p=1,2,3,4$ (from top to bottom) for $N=30$ shells averaged over 
$10^{5}$ realizations.
The continuous lines represent the theoretical scaling
as described in the text. The data for different $p$ are shifted
for a better viewing.
}
\label{fig2}
\end{figure}

In Figure \ref{fig2}a we plot the result of $\langle R^p(t) \rangle^{1/p}$ 
for different sizes of the inertial range.
We indeed observe for $p=2$ a $t^3$ law, while
for higher moments we observe that the nonlinear exponent $\alpha_p$
obtained from (\ref{eq:5}) gives a better fit rather than the linear
scaling $\alpha_p=3p/2$ (which would result in parallel lines).

The scaling exponents satisfy the inequality
$\alpha_p/p < 3/2 $ for $p>2$: this amounts to say
that, as time goes by, the right tail of the pdf
of the separation $R(t)$ becomes less and less broad. 
In other words, due to 
the effect of Eulerian intermittency, particle pairs are more 
likely to stay close to each other than to experience a  large
separation.

\begin{figure}[ht]
\epsfxsize=220pt\epsfysize=183.68pt\epsfbox{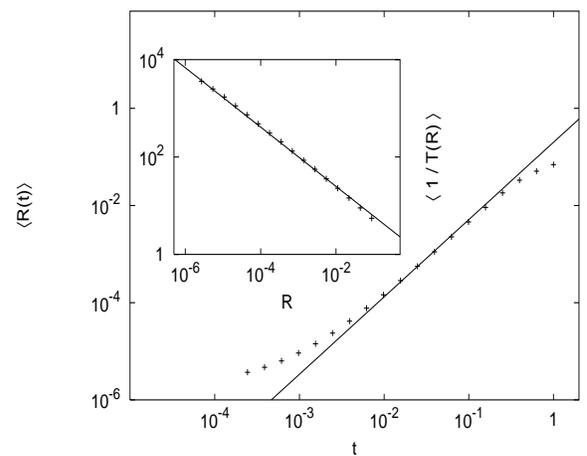}
\caption{Relative dispersion $\langle R(t) \rangle$ for 
for $N=20$ shells averaged over $10^{4}$ realizations.
In the inset we show the corresponding average inverse 
doubling time $\langle 1/T \rangle$. Observe the 
enhancement of the scaling range in the latter case.
}
\label{fig3}
\end{figure}

Figure \ref{fig2}a also shows that the power-law scaling regime for
$\langle R^p(t) \rangle \sim t^{\alpha_p}$ is observed only
well inside the inertial range.
This follows from the dependence of the integration of (\ref{eq:3})
on the smallest scale in the inertial range. 
This effect is particular evident for lower Reynolds numbers,
as shown in figure \ref{fig3} for a simulation with $Re \simeq 10^{6}$.
This correction to a pure power law is far from being negligible for 
instance in experimental data where the inertial range is generally 
limited due to the Reynolds number and the experimental apparatus.
For example, references \cite{FV98,Fung92} show quite clearly the 
difficulties that may arise in numerical simulations with the standard
approach.

Within this framework we propose an alternative approach which is 
based on the statistics at fixed scale, instead of that at fixed time.
The method is in the spirit of a recently introduced generalization
of the Lyapunov exponent to finite size perturbation (Finite Size
Lyapunov Exponent) which has been 
successfully applied in the predictability problem \cite{ABCPV96} 
and in the diffusion problem \cite{ABCCV97}. 
Given a set of thresholds $R_n=R_0 2^{n}$ within the inertial range,
we compute the ``doubling time'' $T(R_n)$ defined as the time it takes 
for the particle separation to grow from one threshold $R_n$ to 
the next one $R_{n+1}$. 
We can give a dimensional estimate of this time as
$T(R) \sim R/\delta v(R)$ and thus see that it fluctuates with the velocity
fluctuations. After averaging over many realizations we can write
\begin{equation}
\left\langle {1 \over T^{p}(R)} \right\rangle \simeq
\int dh R^{p(h-1)} R^{3-D(h)} \simeq R^{\zeta_p-p}
\label{eq:6}
\end{equation}
from which follows that the doubling time statistics contains the same
information on the Eulerian intermittency as the relative dispersion 
exponents (\ref{eq:5}).

As reported in Figure \ref{fig2}b prediction (\ref{eq:6}) is very well 
verified in our simulations. Note that the scaling region is wider 
than that of Figure \ref{fig2}a and the scaling exponent can be
measured with higher accuracy, especially in the case of moderate
inertial range simulation (figure \ref{fig3}).
For this reason we suggest that this kind of analysis should be
preferred when dealing with experimental data.
Also in this case there is an exponent $\zeta_3-3=-2$
unaffected by the intermittency.

In the construction of the synthetic turbulent field we 
have implicitly assumed that the Lagrangian time $\tau_L$ and the
Eulerian time $\tau_E$ are of the same order of magnitude, an assumption
consistent with the experimental data and theoretical arguments
(see, e.g., \cite{McComb}).
Nevertheless it should be interesting to study the relevance of the
ratio $\tau_L/\tau_E$ on the intermittent corrections. One could, indeed,
expects that for $\tau_L/\tau_E\gg 1$ the Lagrangian intermittency 
disappears.

Finally we note that the case of
non intermittent turbulence, i.e., $\zeta_p = p/3$, corresponds to
keep fixed $z_{i,j}=1$. In this case one may ask if
our results are realistic also for the 
Richardson's constant $G_{\Delta}$
defined from the the pair dispersion law
 $R^2(t) = G_{\Delta}\,\overline{\epsilon}\, t^3 $
where $\overline{\epsilon}$ is the average energy dissipation rate. 
The value of $\overline{\epsilon}$ can be obtained from the second order
Eulerian structure function, which reads
$S^{(E)}_2(R)= \langle |\delta v^{(E)}_{\parallel}(R)|^2 \rangle = 
       C_{L}\, \overline{\epsilon}^{2/3}\, R^{2/3}$
where $C_L$ is a universal constant related to the
Kolmogorov constant. According to the experimental measurements 
we fix $C_L=2.0$, leading to 
 $G_{\Delta} = 0.190 \pm 0.005$
for the Richardson's constant which is in agreement with 
previous values \cite{MY75,EM96}.

In this Letter, using a synthetic turbulence model, we have the first
evidence that the relative dispersion statistics for Lagrangian
tracers in fully developed turbulence is affected by the intermittency
of the Eulerian field. Our numerical results are
in agreement with the identification of Lagrangian and Eulerian
intermittency. 
We have suggested a new approach based on the Lagrangian doubling
times which seems very promising for data analysis.
The present work are a first step towards the 
clarification of Lagrangian-Eulerian relationship in fully developed
turbulence. 
It would be extremely interesting to check our claims 
by mean of direct numerical simulations or laboratory experiments.

We thank L. Biferale for useful discussions in the early
stage of the work. This work has been partially supported by 
the INFM (Progetto di Ricerca Avanzata TURBO).


\narrowtext

\end{multicols}
\end{document}